\documentclass[twocolumn,tightenlines,superscriptaddress,eqsecnum,floats,nofootinbib,showpacs]{revtex4-2}
\usepackage{amssymb,amsmath,amssymb,amsfonts,amsthm,stmaryrd,mathrsfs,physics,bbm}
\usepackage{graphicx}
\usepackage{subfigure}
\usepackage[T1]{fontenc}
\usepackage{enumerate} % advanced enumerate environment
\usepackage{color} % for colored text
\usepackage{graphpap} % for numbered coordinate grid
\usepackage{xcolor}

\usepackage{hyperref} % for HyperTeX cross-referencing

\usepackage{amsfonts} 
\usepackage{amsmath}
\usepackage{amssymb}
\usepackage{setspace}
\usepackage{color}

\usepackage{tensor}
\usepackage{tikz}
\usepackage{graphicx,subfigure}% Include figure files
\usepackage[font=small, font+=it, format=hang, margin={1cm, 1cm}]{caption}
\graphicspath{{figure/}}

\usepackage{dcolumn}% Align table columns on decimal point
\usepackage{bm}% bold math

\numberwithin{equation}{section}

\newcommand{\bea}{\begin{eqnarray}}
\newcommand{\eea}{\end{eqnarray}}
\newcommand{\be}{\begin{equation}}
\newcommand{\ee}{\end{equation}}
\def\nn{\nonumber}

\def\p{\partial}

\renewcommand{\d}{\textrm{d}}

\begin{document}

\title{The Repulsive Effect of Covariant Effective Quantum Gravity}
\author{Yan Liu}
\email{yanliu@ytu.edu.cn}
\affiliation{Department of Physics, Yantai University, Yantai 264005, China}

\author{Jie Jiang}
\email{jiejiang@mail.bnu.edu.cn}
\affiliation{School of Physics and Astronomy, Beijing Normal University, Beijing 100875, China}

\author{Bing Sun}
\email{bingsun@bua.edu.cn, corresponding author.}
\affiliation{Department of Basic Courses, Beijing University of Agriculture, Beijing 102206, China}

\begin{abstract}
%In this work we discuss the geodesic motion of covariant effective quantum black hole, and give the explicit critical values of the quantum parameter, beyond which the geodesic orbits disappear. By the analysis of the critical orbital behavior, we find that in covariant effective loop quantum gravity, there exists static free test particle lavitating outside the black hole, and, different with the attractive effect performed by the classical gravity, the quantum correction in the general relativity could perform as the repulsive effect in certain regions.
%As well known, gravity always perform as an attractive effect in general relativity. However, when the quantum effect is considered, this basic intuition may violate and reflect to the motion of test particles. In this letter, by analytically investigating the behavior of the geodesic motion in the recent covariant effective quantum black hole model, we show that gravity could perform as repulsive effect in the Plank scale, especially obvious for those motion with turning points. 
In General Relativity, gravity around a black hole is universally regarded as an attractive force. However, quantum effects may significantly alter this classical picture and thus affect the dynamics of test particles. In this work, we investigate geodesic motion in a recently proposed covariant effective quantum black hole model, the ZLMY type I model in \href{https://journals.aps.org/prd/abstract/10.1103/PhysRevD.111.L081504}{[Phys.Rev.D 111 (2025) 8, L081504]}. Our analytical study reveals that, at the Planck scale, gravity can exhibit a repulsive character, most notably in trajectories involving turning points. These findings suggest that quantum corrections could fundamentally modify our conventional understanding of gravitational interactions in certain regimes.

\end{abstract}

%\pacs{04.65.+e,04.70.-s,11.30.-j,12.10.-g}
%PACS: 04.20.-q, 04.20.Ha, 11.25.Tq, 11.30.Cp

%%%%%%%%%%%%%%%%%%%%%%%%%%%%%%%%%%%%%%%%%%%%%%%%%%%%%%%%%%%%%%%%%%%%%%%%%%%%%%%%%%%%%%%%

\maketitle

\section{Introduction}

In recent years, direct experimental evidence for the predictions of Einstein's General Relativity (GR) has grown significantly. Astrophysical observations, such as the lunar laser ranging \cite{Hofmann:2018myc} and evolution of planetary orbits \cite{Genova:2018myc}, have confirmed the validity of GR in the weak field regime. Meanwhile, the detection of gravitational waves from binary black hole mergers \cite{LIGOScientific:2016aoc} and the first image of a black hole shadow by the Event Horizon Telescope \cite{ET:2019dnz} stand as compelling confirmations of GR's accuracy in the strong field regime. Despite these successes, longstanding theoretical challenges remain, such as the singularity problem and the difficulty of reconciling GR with quantum field theory. These unresolved issues have motivated substantial efforts to develop quantum modifications of gravity. One potential avenue for exploring quantum effects in gravity is to treat quantum gravity as an effective field theory \cite{Donoghue:2012zc}. In this context, the Hamiltonian formulation of classical GR provides a promising framework for investigation. In this formulation, the dynamics of classical GR are governed by a set of first-class constraints: the diffeomorphism and Hamiltonian constraints. To implement effective field theory in the Hamiltonian framework, these constraints must be modified to their effective counterparts. These modified constraints are then interpreted as emergent structures arising from an underlying canonical quantum gravity theory, such as loop quantum gravity  \cite{Rovelli:1987df,Ambjorn:2001cv,Clifton:2011jh,Surya:2019ndm}.

 Since the Hamiltonian formulation depends on the 3+1 decomposition of the spacetime manifold, a key challenge in this framework is restoring full spacetime covariance. Recently, Cong Zhang, Jerzy Lewandowski, Yongge Ma, and Jinsong Yang (ZLMY) \cite{Zhang:2024khj} made significant progress in addressing this issue for spherically symmetric gravity. In their work, the diffeomorphism constraint is retained in its classical form, while the expression for the Hamiltonian constraint is relaxed. By imposing the requirement of general covariance, they derived equations for the effective Hamiltonian constraint. Solving these equations, they proposed several effective Hamiltonian constraints, each corresponding to a covariant effective quantum gravity (EQG) theory. Their analysis provides an ideal setting to explore how quantum effects might modify gravitational properties within a covariant framework.

 A natural starting point for probing the implications of the quantum parameter on spacetime is to examine the geodesic motion of test particles. Studies on geodesic motion are instrumental in understanding astrophysical processes, such as binary inspirals and gravitational wave signals \cite{Yang:2024lmj,Compere:2017hsi,vandeMeent:2017bcc,Poisson:2011nh}, the shadow of black holes \cite{Perlick:2021aok,Hou:2022eev,Lee:2022rtg,Huang:2024wpj}, Penrose process \cite{Shen:2024sdr,Su:2022roo,Penrose:1971uk}, accretion disk \cite{Page:1974he}, and the distribution of dark matter \cite{Sadeghian:2013laa} around compact objects. Recently, considerable attention has been given to the ZLMY models \cite{Zhang:2024khj}, especially through investigations of massive and massless particle motion in these spacetimes \cite{Chen:2025aqh,Lutfuoglu:2025hwh,Al-Badawi:2025yum,Chen:2025ifv,Konoplya:2025hgp,Shu:2024tut,Du:2024ujg,Ban:2024qsa,Heidari:2024bkm,Liu:2024wal,Liu:2024soc,Konoplya:2024lch}. In particular, Du et al. numerically examined the dynamics of spinning test particles \cite{Du:2024ujg}, and discovered that within the ZLMY type I model, the innermost stable circular orbit (ISCO) ceases to exist due to the vanishing angular momentum, when the quantum parameter is sufficiently large. This phenomenon points toward a novel feature in the underlying gravitational interaction. As well known, in GR and some of its modifications, gravity remains strictly attractive for ordinary matter sources, and the conclusion often reinforced by analyzing particle trajectories near black holes \cite{Chen:2024how,Liu:2023tcy,Compere:2021bkk,Druart:2023mba,Cai:2023ygh,Cieslik:2023qdc,Sanchis-Gual:2017bhw,Glampedakis:2002ya,Xamidov:2025oqx}. Nevertheless, under extreme conditions particularly at small scales where quantum effects are expected to dominate, this classical picture may no longer hold \cite{Wei:2019uqg,Lan:2021ngq,Du:2024ujg}. Studying geodesic motion in quantum-corrected spacetimes thus provides a powerful probe of these potential deviations and helps uncover new features of quantum gravity.

In this work, we dive further into the geodesic motion of massive test particles in the covariant EQG black hole spacetime introduced in \cite{Zhang:2024khj}. We first show the disappearance of various timelike geodesic orbits due to the quantum parameter, then discuss the behavior of the critical motion. Our analysis strictly reveals an suppressing phenomenon: beyond a critical value of the quantum parameter, the gravitational field can exhibit effectively repulsive behavior. This stark departure from classical expectations emphasizes the importance of carefully incorporating quantum corrections while preserving covariance. By demonstrating that even well motivated, diffeomorphism covariant quantum modifications can produce repulsive gravitational effects, we raise new questions about the fundamental nature of gravity at the Planckian frontier and beyond.

\section{Review ZLMY type I spacetime}
This work focuses on the first type of static black hole model in \cite{Zhang:2024khj}, referred as the ZLMY type I model in this work. The action of the EQG vacuum spacetime is given by
\bea
S_g=-\int\d t\int\d r(\frac{1}{2}K_1\dot A^1+K_2\dot A^2+N H_{eff}+N^r H_r),\nn\\
\eea
where $(K_2, A^2)$ are the canonical pairs for 2-dimensional gravity, and $(K_1, A^1)$ for the dilaton. $N$ is the lapse function and $N^r$ is the shift vector which are arbitrary spatial fields. The diffeomorphism constraint $H_r$ and the Hamiltonian constraint $H_{eff}$ are given by
\bea
&&H_r=A^2\p_r K_2-K_1\p_r A^1/2,\\
&&H_{eff}=-\frac{A^2}{2\sqrt{A^1}}-\frac{K_1 A^1}{2\alpha}\sin{\frac{2\alpha K_2}{\sqrt{A^1}}}-\frac{3\sqrt{A^1}A^2}{2\alpha^2}\sin^2{\frac{\alpha K_2}{\sqrt{A^1}}}\nn\\
&&+\frac{K_2 A^2}{2\alpha}\sin{\frac{\alpha K_2}{\sqrt{A^1}}}+\frac{(\p_r A^1)^2}{8\sqrt{A^1}A^2}e^{\frac{2i\alpha K_2}{\sqrt{A^1}}}+\frac{\sqrt{A^1}}{2}\p_r(\frac{\p_r A^1}{A^2})e^{\frac{2i\alpha K_2}{\sqrt{A^1}}}\nn\\
&&+\frac{i\alpha A^2}{4}(\frac{\p_r A^1}{A^2})^2(\frac{K_1}{A^2}-\frac{K_2}{A^1})e^{\frac{2i\alpha K_2}{\sqrt{A^1}}}.
\eea

By varying the quantities $N$, $N^r$, $K^I$, and $A^I$ in the action, the modified Einstein field equations (EFE) can be obtained. Because the explicite form of these equations is lengthy and complicated, we refer the readers to Ref.~\cite{Zhang:2024khj} for more details. Solving the modified EFE under the gauge conditions $A^1=r^2$ and $N^r=0$, the expression of the ZLMY type I metric is obtained which is given by
%By choosing the gauge $A^1=r^2$, fixing $N$ and $N^r$ with the stationary equations $\{A^I(r), H_{eff}[N]+H_r[N^r] \}=0$ for $I=1,2$, and choosing $N^r=0$ to fix the gauge corresponding to the effective Hamiltonian constant, the expression of the ZLMY type I metric can be solved which is given by 

\bea
&&\d s^2=-f(r)\d t^2+\frac{\d r^2}{f(r)}+r^2\d \theta^2+r^2\sin^2\theta\d\phi^2,\\
&&f(r)=1-\frac{2M}{r}+\frac{\alpha^2}{r^2}(1-\frac{2M}{r})^2,\label{fr}
\eea
where $M$ is the mass of the covariant black holes, and $\alpha$ is the quantum parameter. The outer horizon and inner horizon of the black hole locate at $r_+$ and $r_-$ which are given by 
\bea
&&r_+=2M, \quad r_-=\frac{A}{3}-\frac{\alpha^2}{A},\\
&&A^3=3\alpha^2(9M+\sqrt{81M^2+3\alpha^2}).
\eea

%The outer event horizons $r_{1+}=r_{2+}=2M$ are independent of the quantum parameter. While the inner horizons are given by 
%\bea
%&&r_{1-}=r_{2-}=\frac{A}{3}-\frac{\alpha^2}{A},\\
%&&A^3=3\alpha^2(9M+\sqrt{81M^2+3\alpha^2}).
%\eea
% As the parameter $\alpha$ increases to infinity, the inner horizon asymptotically close to the outer horizon, which indicates that when the mass of the black hole $M$ is small as in the quantum scale level, the black hole is nearly extreme. When $\alpha=0$, the inner horizon vanishes, and turns out to be the physical singularity at $r=0$.

 Since spacetime is spherically symmetric, we consider the motion on the equatorial plane such that $\theta=\pi/2$ and $\dot\theta=0$, where the dot over $\theta$ represents the derivative with respect to proper time $\tau$. In the equatorial plane, the behavior of geodesics is dominated by the radial motion $\dot r=\pm \sqrt{R(r)}$, where the signs $+$ and $-$ denote the ingoing and outgoing trajectories, and the radial potential $R(r)$ is given by
\bea
R(r)&=&E^2-f(r)(1+\frac{L^2}{r^2}),
\eea
where $E$ and $L$ are the conserved energy and orbital angular momentum of the particle per unit mass. For simplicity, in the following sections we set $M=1$ in the figures.

%We can rewrite $R_2(r)$ as 
%\bea
%&&R_2(r)=\frac{\mu_2(r)}{r^3}V_{eff},\\
%&&V_{eff}=L_2^2 (2 M - r) + r^2 (2 M + (-1 + E_2^2) r),
%\eea

\section{Circular orbits}\label{sec3}
For generic circular geodesic motion, the radial position and velocity satisfy $r_*=\mathrm{Const.}$, $\dot{r}_*=0$ or equivalently the radial potential satisfies $R(r_*)=0$ and $\frac{\d R(r_*)}{\d r}=0$. Explicitly, the two equations are
\begin{equation}
\begin{aligned}
     &E^2-f(r_*) (1+\frac{L^2}{r_*^2}) = 0\,, \\
     &\beta (1+\frac{L^2}{r_*^2}) - L^2r_*^2f(r_*) = 0\,,
\end{aligned} \label{rdotr}
\end{equation}
where $\beta$ is given by
\begin{equation}
    \beta = Mr_*^3 - \alpha^2(r_* - 2M)(r_* - 4M)\,.\label{noISCO}
\end{equation}
By solving equations \eqref{rdotr}, the conserved energy and angular momentum of the circular motion can be obtained given by
\bea
&&E(r_*,\alpha)=\frac{(r_*-2M)(r_*^3+\alpha^2(r_*-2M))}{r_*^2\sqrt{(r_*-3M)(r_*^3+2\alpha^2(r_*-2M))}}\label{Ecir}\,,\\
&&L(r_*,\alpha)=\sqrt{\frac{r_*^2\beta}{(r_*-3M)(r_*^3+2\alpha^2(r_*-2M))}}\,.\label{Lcir}
\eea
 With these expressions for $(E, L)$ as functions of $r_*$, the allowed domain for circular motion is determined by requiring the arguments of the square roots in \eqref{Ecir} and \eqref{Lcir} to be non-negative, which ensures the orbital energy and angular momentum are real and therefore are physical meaningful.

As is well known in Schwarzschild spacetime, the innermost stable circular motion (ISCO) is located at $r_{ISCO}=6M$, while unstable circular motion is allowed between the innermost circular orbit (ICO) $r_{ICO}=3M$ and $r_{ISCO}=6M$, the stable circular motion exists beyond the ISCO. In this quantum corrected case, based on the analysis of the positive property of the formula within the square root of $L$, the position of the ICO remains fixed at $r_{ICO}=3M$, as determined by the denominator factor $r_*-3M$ in \eqref{Lcir}, which are independent of the quantum correction $\alpha$. However, unlike the Schwarzschild case, circular motions are not always permitted for all $r_*>r_{ICO}$. The quantum correction $\alpha$ introduces new constraints, the reality condition for the angular momentum $L$ in \eqref{Lcir}, i.e. the positive property of $\beta$.

 In the Schwarzschild limit, one finds $\beta=Mr_*^3$, which is strictly positive for $r_*>0$. Therefore, the angular momentum of the circular motion is always real in the region $r_*>r_{ICO}$, ensuring that the circular motion are always allowed outside the ICO. However, in the ZLMY type I model, the positive property of $L^2$ \eqref{Lcir} is decided by the non-trivial cubic polynomial numerator $\beta$ \eqref{noISCO}, whose sign depends on the value of $\alpha$. As a result, there exists a critical value of the parameter $\alpha=\alpha_c$, beyond which $\beta$ becomes negative in certain radial regions. This implies that $L^2$ becomes negative and circular motion is forbidden there.

The explicit range of the "no circular" region can be obtained by finding the root of the cubic polynomial $\beta$\eqref{noISCO}. The critical value of the quantum parameter and the circular motion is given by
\bea
\alpha_c=2\times 3^{3/4}M, \qquad r_c=2(\sqrt{3}+3)M.
\eea
Note that the circular motion is always allowed for $\alpha\leq\alpha_c$, since $\beta$ keeps positive outside of the horizon. With $\alpha>\alpha_c$, there exists three distinct real roots for $\beta$ given by\\
%When $\alpha>2\times3^{3/4}M$, we have $\Delta_3>0$, there exists three distinct real roots for $\beta$ given by

%The discriminate of $\beta$ is given by 
%\bea
%\Delta_3=\frac{4\alpha^4(\alpha^4-432M^4)}{M^2}.
%\eea

%When $\alpha\leq2\times3^{3/4}M$, we have $\Delta_3\leq0$, and outside of the horizon, we always have $\beta\geq 0$, such that the circular motion is always allowed.\\
%When $\alpha>2\times3^{3/4}M$, we have $\Delta_3>0$, there exists three distinct real roots for $\beta$ given by
\bea
r_*^{(i)}&=&\frac{1}{3}\left(\alpha^2+2\alpha\sqrt{\alpha^2-18M^2}\cos(\zeta+\frac{2\pi i}{3})\right),\\
\zeta &=&\frac{1}{3}\arccos(\frac{108M^4-27M^2\alpha^2+\alpha^4}{\alpha\sqrt{(\alpha^2-18M^2)^3}}),\quad i=0,1,2.\nonumber\\
\label{ri}
\eea

\begin{figure}
    \centering
    \includegraphics[width=0.9\linewidth]{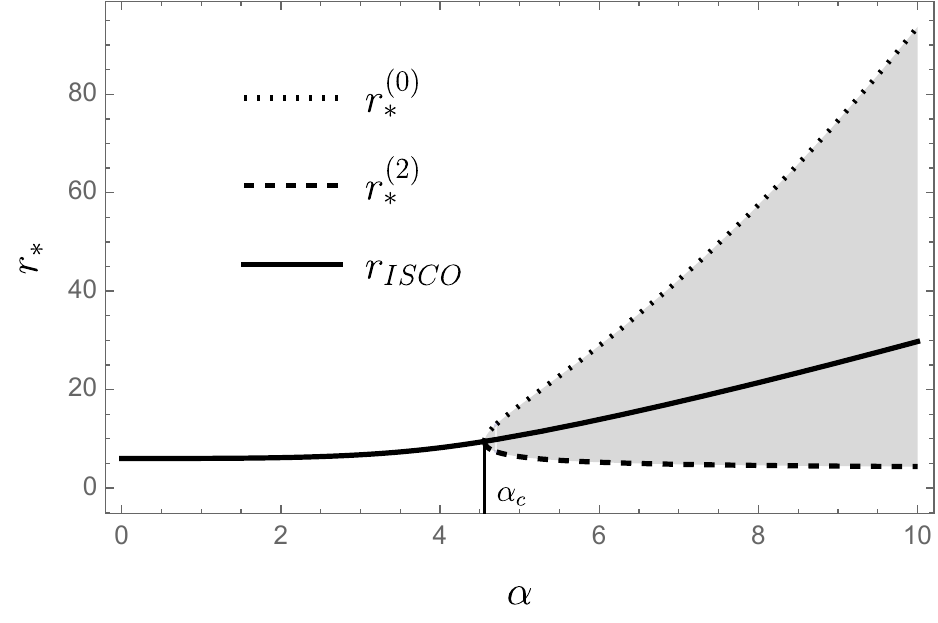}
    \caption{The disallowed (gray) region for circular motion.}
    \label{fig:rISCO1}
\end{figure}

Note that $r_*^{(0)}>r_*^{(2)}>r_+>r_*^{(1)}$, and when $\alpha=\alpha_c$, the roots $r_*^{(0)}$ and $r_*^{(2)}$ merge into a double root to be $r_c$. In the range $r_*>r_*^{(0)}$ and $r_+<r_*<r_*^{(2)}$, the polynomial $\beta>0$, which implies that the circular motion is allowed. In the range $r_*^{(2)}<r_*<r_*^{(0)}$, the polynomial $\beta<0$, the angular momentum of the circular motion is unreal, which implies that the circular motion is disallowed. Therefore, when $\alpha>\alpha_c$, the forbidden region for circular motion starts to appear, as shown in Figure \ref{fig:rISCO1}.

%The position and conserved energy and angular momentum of the ISCO motion can be obtained by solving $\frac{\d^2 R(r)}{dr^2}=0$ combined with the solutions \eqref{Ecir}. 
In Figure \ref{fig:rISCO1}, we show the disappearance of the circular motion, including the vanished ISCO. The dotted line and dashed line denote the boundary of the "no circular" region corresponding to $r_*^{(0)}$ and $r_*^{(2)}$ respectively. Between these two boundaries (in the gray-shaded area), circular motion is prohibited. The black line $r_{ISCO}$ traces the position of $r_{ISCO}$ as a function of $\alpha$, intersecting with $r_*^{(0)}$ and $r_*^{(2)}$ at $\alpha=\alpha_c$. It is important to note that in the region $r_*^{(2)} < r < r_{ISCO}$, the unstable circular orbits are forbidden, and in the region $r_{ISCO}<r<r_*^{(0)}$, the stable circular orbits become disallowed. Figure \ref{fig:rISCO1} demonstrates that quantum correction causes the ISCO to shift significantly from its classical position. When $\alpha>\alpha_c$, the ISCO disappears first, and, as $\alpha$ increases further, both the stable and unstable circular orbits disappear. In addition, in the asymptotic limit $\alpha\to\infty$, we have $r_*^{(2)}\to 4M$.

As shown in Figure \ref{fig:CIrEL}, the marginal circular orbits (MCO) corresponding to $E=1$ separate the bound and unbound motion in the phase space of $(E,L)$. The angular momentum aligns with that in equations \eqref{Lcir}. In turn, the radius of MCO $r_m$ is the root of an eighth-order polynomial expressed as \eqref{rm}. Combined with $r_m=r_*^{(2)}$, the critical values of MCO are given by
\bea
\alpha_m=3\sqrt{3}M, \qquad r_{mc}=6M.
\eea
Note that as $\alpha$ increases, the angular momentum of ISCO and MCO monotonically decrease to zero at $\alpha=\alpha_c$ and $\alpha=\alpha_m$. This implies that at this point, although the radial potential still satisfies zero conditions, including its first derivative, the ISCO and MCO do not behave as a circular motion, but as a rest particle hanging at $r_c$ and $r_{mc}$, since their angular momentum are zero. Similarly for those orbits with $\alpha>\alpha_c$, the critical circular orbits turn out to be a rest particle lavitating at $r_*^{(0)}$ or $r_*^{(2)}$. The difference between these two points is that, at $r_*^{(0)}$, the rest particle is stable, while at $r_*^{(2)}$ it is unstable.

 It is worth emphasizing that along the curves $r_*^{(0)}$ and $r_*^{(2)}$ in Figure \ref{fig:rISCO1}, the function $\beta$ vanishes, leading to zero angular momentum $L=0$. Despite this, these motions are still permitted because they satisfy the circular conditions $R(r_*)=0$ and $\d R(r_*)/\d r_*=0$, with both the orbital angular momentum and energy remaining real. However, such critical motions do not behave like conventional circular orbits. Instead, they describe particles effectively "levitating" at a fixed radial position $r_*$, without undergoing orbital and azimuthal motion. This phenomenon arises purely from the modified gravitational structure of the ZLMY type I spacetime, and can be interpreted as a manifestation of an exact balance between gravitational attraction and a repulsive component induced by quantum corrections. Therefore, it highlights a remarkable feature of gravity in this model, the emergence of repulsive behavior near the black hole. This repulsive nature becomes even more evident in the analysis of geodesics with turning points, such as bound or deflecting trajectories, which will be discussed in the subsequent sections.

\section{Motion with turning points}\label{sec4}
\begin{figure}
    \centering
    \includegraphics[width=0.9\linewidth]{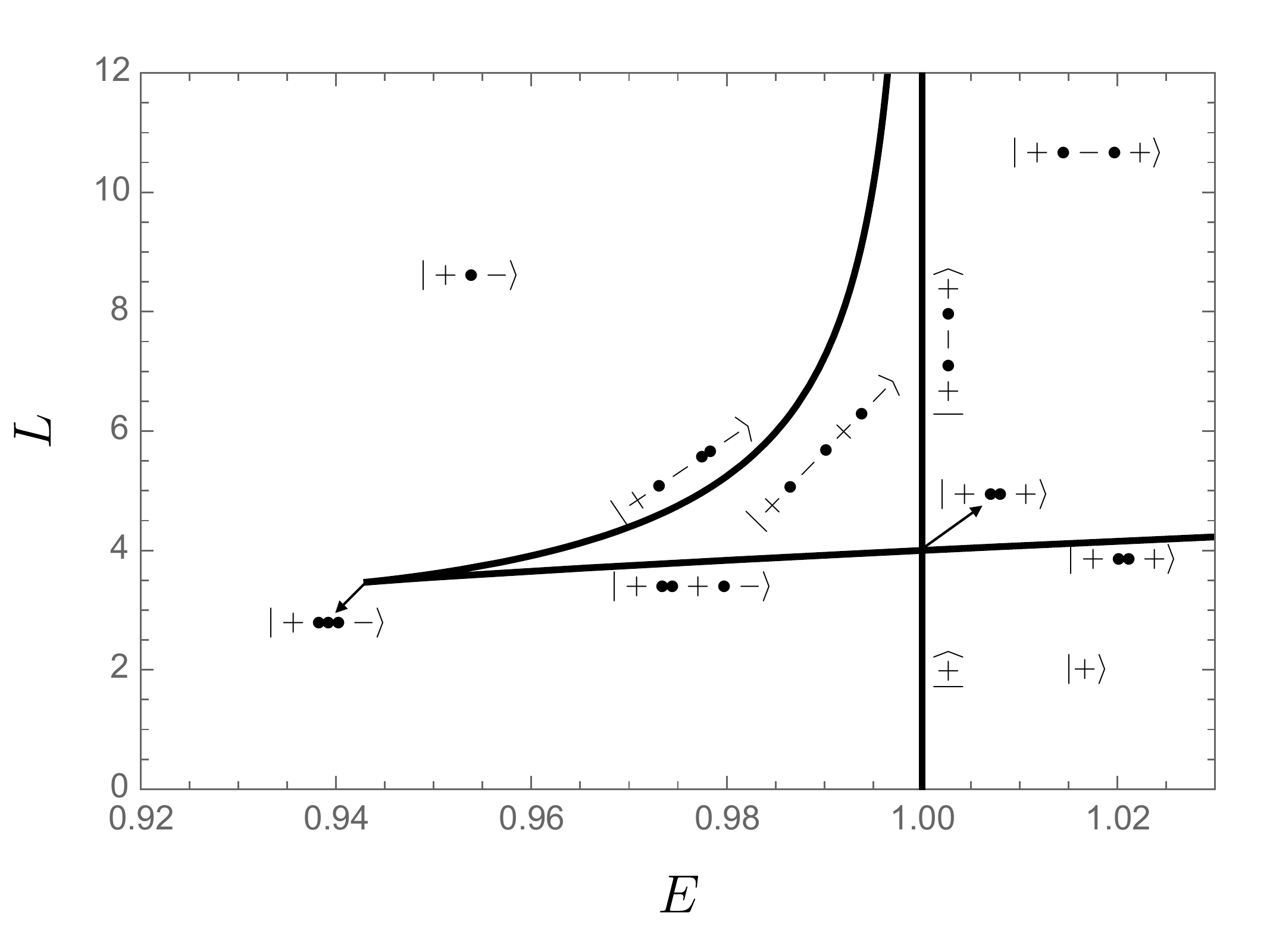}
    \caption{The classification of Schwarzschild geodesic motion in $(E,L)$ phase space.}
    \label{fig:SchwarzEL}
\end{figure}

 In the previous section, we demonstrated the vanishing of circular orbits in the phase space of $(E,L)$. Based on the continuity of the geodesic motion in the phase space, circular orbits correspond to double roots of the radial potential, and in their vicinity, such a degenerate root bifurcates into two distinct single roots that demarcate turning points, thus organizing the orbital structure into trapped, bound, and deflected trajectories as classified in \cite{Compere:2021bkk}. Therefore, it is natural to deduce that there exists a region such that the generic trapped, plunging, bound, and deflecting orbits disappear. To facilitate systematic analysis, Table \ref{table:notations} summarizes the root structure notation for radial potentials, while Figure \ref{fig:SchwarzEL}, adapted from \cite{Compere:2021bkk}, illustrates the phase space classification of geodesics in Schwarzschild spacetime. This notation covers the complete root structure of the radial potential $R(r)$ from the event horizon to spatial infinity, providing a convenient framework for discussion. 
 
 For example, consider the root structure $\vert+\bullet-\bullet+\bullet-\rangle$, moving outward from the horizon $\vert$ to infinity $\rangle$, $R(r)$ possesses three single roots denoted by $\bullet$, corresponding to the turning points $r_1$,  $r_2$ and $r_3$. The first $+$ sign indicates that $R(r)>0$ in the region $r_+<r<r_1$, where geodesic motion is permitted, while the subsequent $-$ shows that $R(r)<0$ in the region $r_1<r<r_2$, where the geodesic motion is not allowed. Thus, motion in the first $+$ region are confined in $r_+<r<r_1$, whereas the second $+$ represents the periodic bound orbits between $r_2$ and $r_3$. As the orbital energy and angular momentum vary in the phase space, $r_2$ and $r_3$ can coalesce into a double root, forming a stable circular orbit denoted by $-\bullet\hspace{-3pt}\bullet-$. Then the overall root structure becomes $\vert+\bullet-\bullet\hspace{-3pt}\bullet-\rangle$.

%\begin{table}[!tbh]    \centering
%\begin{tabular}{|c|c|c|c|c|}\cline{1-2}\cline{4-5}
%\rule{0pt}{13pt}\textbf{Notation} & \textbf{Denotes} & & \textbf{Notation} & \textbf{Denotes}\\\cline{1-2}\cline{4-5}
%$\vert$ &  outer horizon & & $\bullet$ & simple roots (turning points) \\\cline{1-2}\cline{4-5}
%$+$ & allowed region & &  $\bullet \hspace{-2pt}\bullet$ & double roots (circular orbits)\\\cline{1-2}\cline{4-5}
%$-$ & disallowed region & & $\bullet\hspace{-4pt}\bullet\hspace{-4pt}\bullet$ & triple roots (ISCO)\\\cline{1-2}\cline{4-5}
%$\rangle$ & radial infinity & & ${\vert \hspace{-5pt} \bullet }$ & roots touching the horizon\\\cline{1-2}\cline{4-5}
%\end{tabular}\caption{Notations for the root structures of the radial potential.}\label{table:notations}
%\end{table}

\begin{table}[!tbh]    \centering
\begin{tabular}{|c|c|}\cline{1-2}
\rule{0pt}{13pt}\textbf{Notation} & \textbf{Denotes} \\\cline{1-2}  
$\vert$ &  outer horizon  \\\cline{1-2}
$+$ & allowed region \\\cline{1-2}
$-$ & disallowed region \\\cline{1-2}
$\rangle$ & radial infinity \\\cline{1-2}
 $\bullet$ & simple roots (turning points) \\\cline{1-2}
  $\bullet \hspace{-2pt}\bullet$ & double roots (circular orbits)\\\cline{1-2}
 $\bullet\hspace{-4pt}\bullet\hspace{-4pt}\bullet$ & triple roots (ISCO)\\\cline{1-2}
\end{tabular}\caption{Notations for the root structures of the radial potential.}\label{table:notations}
\end{table}

\begin{figure}
    \centering
    \includegraphics[width=0.9\linewidth]{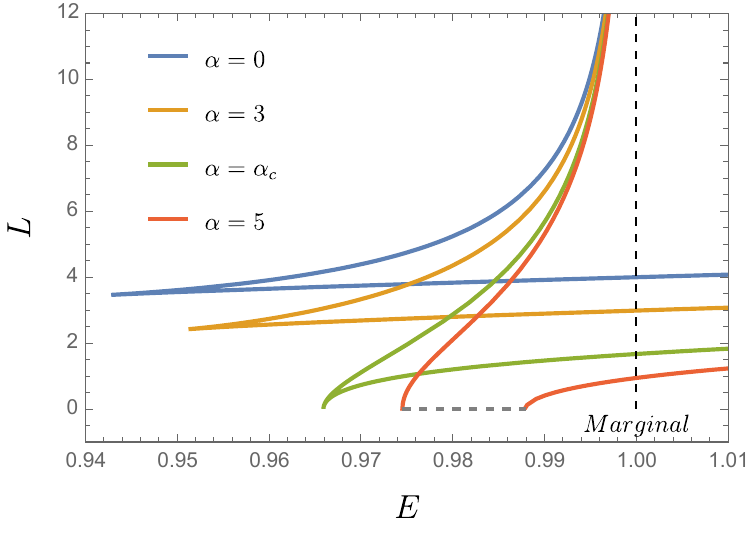}
    \caption{The $(E,L)$ phase space of the geodesic motion. The colored curves denote the circular motion.}
    \label{fig:CIrEL}
\end{figure}

Figure~\ref{fig:CIrEL} illustrates how the phase space structure varies with the parameter $\alpha$. When $\alpha = \alpha_c$, the ISCO begins to vanish and for $\alpha > \alpha_c$, a forbidden region emerges in which circular motion is no longer allowed. This results in the elimination of both stable and unstable circular orbits, as indicated by the thick gray dashed line in Figure~\ref{fig:CIrEL}. The root structures associated with stable and unstable circular orbits with $E < 1$, $\vert+\bullet-\bullet\hspace{-3pt}\bullet-\rangle$, and $\vert+\bullet\hspace{-3pt}\bullet+\bullet-\rangle$, contain not only the circular motion denoted by $\bullet\hspace{-2pt}\bullet$, but also a variety of non-circular motions within the $+$ region. These structures include trapped orbits of the form $\vert+\bullet\hspace{2pt}-$, whirling trapped orbits represented by $\vert+\bullet\hspace{-2pt}\bullet$, and homoclinic orbits described by $\bullet\hspace{-3.5pt}\bullet+\hspace{2pt}\bullet$. All of these orbits share the same energy and angular momentum as the corresponding circular orbits in phase space. Consequently, when circular orbits are eliminated due to the emergence of the forbidden region, these related orbit types also disappear.

The vertical black dashed line in Figure~\ref{fig:CIrEL} represents the marginal motion with energy $E = 1$, and its intersections with the colored curves correspond to marginally circular orbits. Notably, when $\alpha > \alpha_{m}$, the gray dashed line lies above the threshold $E = 1$  and intersects the black marginal line, indicating that the MCOs, along with portions of the unstable circular orbits with $E > 1$, characterized by the root structure $\vert+\bullet\hspace{-3pt}\bullet+\rangle$, vanish. As discussed in the previous paragraph, the associated whirling trapped orbits and whirling deflecting orbits $\bullet\hspace{-4pt}\bullet+\rangle$ are also prohibited in this regime.

As shown in Figure~\ref{fig:SchwarzEL}, periodic bound motion is confined within a triangular region bounded by circular and marginal orbits, characterized by the root structure $\vert+\bullet-\bullet+\bullet-\rangle$. By continuity in phase space, the turning points of the bound orbits with structure $\bullet+\bullet$ can be interpreted as a separation of the double root $-\bullet\hspace{-4pt}\bullet-$ from the stable circular orbit or of $\bullet\hspace{-4pt}\bullet+\bullet$ from the unstable circular orbit. Therefore, it follows naturally that when circular orbits vanish, the neighboring bound orbits in phase space also disappear.
As illustrated by the red curve with $\alpha > \alpha_c$ in Figure~\ref{fig:CIrEL}, one can visualize a missing triangle-shaped horn region excised by the thick gray dashed line. This boundary corresponds to the vanished circular orbits, and the interior of this region contains the disappeared bound orbits. In particular, the thick gray dashed line itself represents the critical bound motion with vanishing angular momentum, $L = 0$.
Similarly, when $\alpha > \alpha_{mc}$, the disappearance of the unstable circular orbit $\vert+\bullet\hspace{-3pt}\bullet+\rangle$ leads to the elimination of the deflecting orbits in the second $+$ region of the root structure $\vert+\bullet-\bullet+\rangle$. In the following analysis, we focus on the nonexistence of bound orbits as a representative case, and subsequently extend the discussion to deflecting orbits.

The turning points of bound motion are the pericenter $r_p = \frac{p}{1 + e}$ and the apocenter $r_a = \frac{p}{1 - e}$, where $e$ and $p$ denote the eccentricity and semi-latus rectum of the orbit, respectively. The energy and angular momentum associated with the bound motion can be determined by solving the conditions $R(r_a) = R(r_p) = 0$. By analyzing the reality condition of the angular momentum, the critical value $\alpha_{bc}$ for the bound motion can be obtained and is given by
\bea
\alpha_{bc} = \sqrt{\frac{M p^3}{(4M - p)\left[2M(1 + e^2) - p\right]}}.
\eea

%\bea
%R(r_a)=R(r_p)=0,\label{Rbound}
%\eea
%replacing $r_a$ and $r_p$ by the taxonomy of eccentricity $e$ and semi-latus rectum $p$ as
%\bea
%r_a=\frac{p}{1-e},\qquad r_p=\frac{p}{1+e},
%\eea
%and solve \eqref{Rbound}, the energy and angular momentum of the bound motion are obtained given by
%\bea
%E&=\frac{1}{p^3}((2 (1 + e) M - p) ((1 + e)^2 L^2 + p^2)\nonumber\\
%&(  (2 (1 + e) M - p)(1 + e)^2 \alpha^2-p^3 ))^{\frac{1}{2}},\\
%L&=(\frac{p^2(M p^3 - (4 M - p) (2 (1 + e^2) M - p) \alpha^2)}{(p-((3 + e^2) M) ) (p^3 + 
%   2 (p + e^2 p-2 M - 6 e^2 M) \alpha^2)})^{\frac{1}{2}}.
%\eea
%\bea
%E^2&=&\frac{1}{p^6}(2 (1 + e) M - p) ((1 + e)^2 L^2 + p^2)\nonumber\\
%&&(  (2 (1 + e) M - p)(1 + e)^2 \alpha^2-p^3 ),\\
%L^2&=&\frac{p^2(M p^3 - (4 M - p) (2 (1 + e^2) M - p) \alpha^2)}{(p-((3 + e^2) M) ) (p^3 + 
   %2 (p + e^2 p-2 M - 6 e^2 M) \alpha^2)}.\nonumber\\
%\eea
%The angular momentum of bound orbits are real when $\alpha<\alpha_{bc}$, where the critical parameter $\alpha_{bc}$ for the bound motion is given by
%\bea
%\alpha_{bc}=\sqrt{\frac{Mp^3}{(4 M - p) (2M (1 + e^2)- p)}}.
%\eea

\begin{figure}
    \centering
    \includegraphics[width=0.9\linewidth]{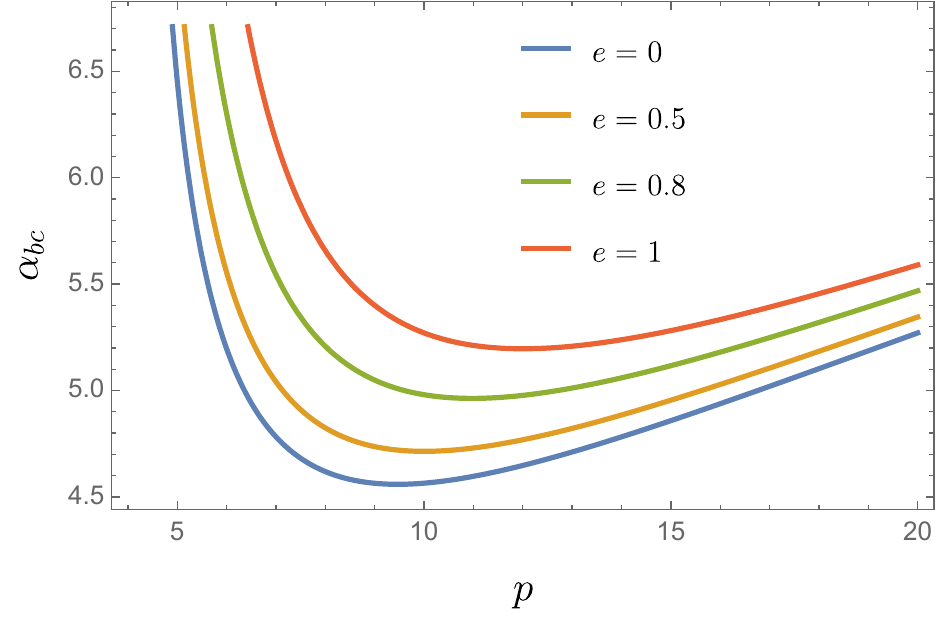}
    \caption{The critical value $\alpha_{bc}$ in terms of $e$ and $p$ for bound orbital motion.}
    \label{fig:abc}
\end{figure}

In Figure~\ref{fig:abc}, we show the critical value $\alpha_{bc}$ as a function of eccentricity $e$ and semi-latus rectum $p$. In the case of $e = 0$, the two turning points in the root structure $-\bullet+\bullet-$ merge into a double root $-\bullet\hspace{-4pt}\bullet-$, corresponding to $r_a = r_p = r_* = p$, where $p$ denotes the radius of the critical stable circular orbit. The minimum value of the curve in this case coincides precisely with $\alpha_c$.
As eccentricity increases in the range $0 < e < 1$, such as in the cases $e = 0.5$ and $e = 0.8$, the critical value $\alpha_{bc}$ increases, denoting the start point of the disappearance of bound orbits. For these values of $e$, the minimum point of each curve corresponds to the disappearance of unstable circular orbits and homoclinic trajectories characterized by the root structure $\bullet\hspace{-3.2pt}\bullet+\bullet$. At these points, the pericenter $r_p$ merges with the first turning point of the trapped orbits, forming a double root that represents the unstable circular orbit. The region between this double root and the apocenter contains homoclinic orbits.
The case $e = 1$ describes the disappearance of the marginal motion with energy $E = 1$, and the minimum value of the curve is precisely the critical parameter $\alpha_{m}$ for the MCO. In this scenario, the apocenter tends to infinity, while the pericenter coincides with the turning point of the trapped orbit, forming an unstable MCO. The associated root structure, including whirling trapped and whirling deflecting orbits, is $\vert+\bullet\hspace{-3pt}\bullet+\rangle$. At other points along the $e = 1$ curve, the double root splits into distinct pericenter and turning points, producing the root structure $\vert+\bullet-\bullet+\rangle$ composed of trapped orbits and marginal deflecting orbits.
When $e > 1$, the apocenter becomes negative; however, the critical parameter $\alpha_{bc}$ is still given by Eq.~\eqref{fig:abc}, and its qualitative behavior remains similar to that shown in Figure~\ref{fig:abc}, beyond the boundary $e = 1$. The polar (minimum) points of the curves for $e > 1$ correspond to the critical values at which unstable circular orbits with $E > 1$ disappear, along with the associated whirling trapped and whirling deflecting orbits in the root structure $\vert+\bullet\hspace{-4pt}\bullet+\rangle$. Other points along these curves indicate the disappearance of deflecting orbits characterized by the root structure $\vert+\bullet-\bullet+\rangle$.

\section{Analysis on the critical behavior}
In classical gravity, aspects on the geodesic motion have been widely studied.  Typically, bound orbits possess non-zero angular momentum, implying that the ingoing and outgoing trajectories ---separated by a turning point --- do not follow the same path. Instead, the particle winds around the black hole several times, a motion characterized by well-known rational numbers~\cite{Levin:2008mq}.
The whirling deflecting orbits with $E \geq 1$ in the $+$ region of the root structure $\bullet\hspace{-4pt}\bullet+\rangle$ describe particles that originate from spatial infinity with finely tuned energy and angular momentum. These trajectories spiral inward and asymptotically approach an unstable circular orbit around the black hole. Similarly, the homoclinic orbits in the $+$ region of $\bullet\hspace{-3pt}\bullet+\bullet-\rangle$ with $E < 1$ also asymptotically approach an unstable circular orbit, but instead of starting from infinity, the motion begins at a point between the apocenter and the unstable orbit.
The deflecting orbits in the $+$ region of the root structure $-\bullet+\rangle$ with $E > 1$ also originate from infinity and eventually escape back to infinity. During their trajectory, the particle may move around the black hole several times before reaching the pericenter, and then circle it several more times as it flies away. All of these generic classes of motion typically involve several circles around the black hole. Even when a full orbit is not completed, they still result in a non-zero azimuthal change $\Delta\phi$ due to the presence of non-zero angular momentum.
Especially, when a particle reaches a turning point, it does not simply bounce back along its original trajectory, and instead it departs along a different path.

However, in effective quantum gravity, geodesic motion exhibits notable differences, especially in cases involving critical parameter values. For trajectories corresponding exactly to the critical quantum parameter, bound orbits can still exist with vanished angular momentum $L = 0$, implying an intriguing phenomenon: effective quantum gravity not only exerts an attractive force, but also exhibits a repulsive component.
As discussed in Section~\ref{sec3}, at the critical point, instead of undergoing circular-like orbital motion, the particle remains static, levitating at $r_*$. This behavior provides preliminary evidence for a repulsive component in the effective quantum gravitational interaction. Moreover, in both bound and deflecting motion scenarios, the repulsive effect becomes even more pronounced.

For the bound orbits at the critical value $\alpha = \alpha_{bc}$, one can consider a free-fall particle, with energy and angular momentum properly selected in phase space, initially placed at any point $r_b$ between the pericenter $r_p$ and the apocenter $r_a$. The radial potential vanishes at the turning points, i.e., $R(r_a) = R(r_p) = 0$, and remains positive in the interval $r_p < r < r_a$. This implies the existence of a turning point of the potential, denoted $r_o$, such that $R'(r_o) = 0$ and $r_p < r_o < r_a$.
Due to the vanishing angular momentum, the particle falls radially inward to the pericenter and then reverses direction, returning to the apocenter along the same path. As a result, the entire bound motion behaves analogously to a one-dimensional harmonic oscillator, bouncing back and forth between $r_a$ and $r_p$ without orbiting the black hole.
This periodic bound motion clearly indicates that gravity is attractive in the region $r_o < r < r_a$, where the particle accelerates inward, and repulsive in the region $r_p < r < r_o$, where it decelerates and reverses direction.

For critical homoclinic orbits with $E < 1$, which extend from the apocenter $r_a$ to the double root $r_*$, the freely falling particle moves radially inward toward $r_*$ and ultimately stops at the critical unstable circular position $r_*$. Similarly, for critical whirling deflecting orbits with $E \geq 1$, the particle falls from spatial infinity and asymptotically approaches the same unstable circular orbit at $r_*$, where it hangs at the critical unstable circular position in the end.

For critical deflecting orbits with $E \geq 1$, the particle may fall freely from spatial infinity, move directly to the turning point, and then rebound back to infinity along precisely the same ingoing path. This type of geodesic motion clearly manifests the repulsive effect present in covariant effective quantum gravity.
It is worth noting that there still exists a polar point $r_p < r_o < \infty$ for these critical deflecting orbits. In the region $r_o < r < \infty$, the gravitational interaction is attractive, drawing the particle inward. In contrast, within the region $r_p < r < r_o$, the particle experiences a repulsive force, which causes it to reverse direction at the turning point.

 In summary, while classical geodesic motion around black holes typically involves nonzero angular momentum and multi-winding trajectories that reflect the purely attractive nature of gravity, the effective quantum gravity framework introduces novel behaviors. At critical values of the quantum parameter, the appearance of circular, bound and deflecting orbits with vanishing angular momentum marks a qualitative departure from classical intuition. The existence of such orbits, where particles can levitate statically at $r_*$ or undergo purely radial periodic motion, implies a repulsive component inherent to the quantum corrected spacetime. This repulsive effect is especially evident in the reversal of motion at turning points along the same radial trajectory, even in the absence of angular momentum. Thus, effective quantum gravity modifies the geodesic structure not merely quantitatively but qualitatively, encoding both attractive and repulsive features that have no classical analogue.

\section{Conclusion and discussion}
In this work, we have investigated the behavior of geodesic motion in the type I ZLMY spacetime. We found that there exist critical values of the quantum correction parameter, beyond which all classes of geodesic motion in the $(E, L)$ phase space start to disappear. These include the ISCO, generic circular orbits, periodic bound orbits, deflecting trajectories, and other motion types related to them.

We analytically examined the critical behavior of various types of geodesic motion and provided an explicit determination of the corresponding critical parameters. Our results revealed that the critical circular geodesic motion deviated from the conventional sense of circular motion, instead resembling a static point suspended in space. This finding suggested the existence of static, free test particles around the quantum black holes within the framework of covariant effective quantum gravity.

We compared the behavior of geodesic motion with turning points in Schwarzschild spacetime and in ZLMY type I spacetime. In the latter case, due to the vanishing angular momentum, the critical periodic bound motion transitions into a purely radial oscillation between the pericenter and apocenter. Similarly, the critical homoclinic and whirling deflecting orbits asymptotically approach and settle at their respective unstable circular positions, as determined by their root structures.
Most notably, the behavior of critical deflecting orbits provides clear evidence that covariant effective quantum gravity introduces a repulsive component to the gravitational interaction, supplementing the familiar attractive force.

The emergence of the repulsive gravitational regime in ZLMY type I spacetime carries significant implications for our understanding of gravity near the Planck scale. Traditionally, gravitational attraction is regarded as a universal property, with orbits and trajectories around black holes governed solely by attractive potentials. However, the appearance of static free test particles and repulsion-like behavior challenge this classical intuition, suggesting that quantum effects may profoundly alter gravitational dynamics.
Especially, when the quantum correction parameter becomes dominant, the characteristic size of the black hole approaches the Planck scale. In this regime, quantum gravitational effects dominate and can give rise to such counterintuitive phenomena.

It is worth emphasizing that the repulsive behavior of quantum-scale black holes is not unique to our scenario. Within a thermodynamic framework, similar repulsive interactions have been observed in regular and Anti–de Sitter (AdS) small black holes~\cite{Wei:2019uqg,Lan:2021ngq}. In those contexts, the repulsive effect arises cause of the violation of the strong energy condition, where the Einstein tensor is treated as an effective stress-energy tensor.
However, within the framework of quantum gravity, the non-vanishing Einstein tensor originates from quantum effects rather than from classical matter fields. This approach effectively interprets quantum corrections as classical source terms, which may obscure or neglect essential quantum properties of the system.
However, several open questions remain. While our analysis reveals the emergence of repulsive interactions within a classical dynamical framework, a fully quantum treatment of this phenomenon has yet to be developed. From an experimental perspective, detecting quantum-scale black holes poses significant challenges, especially since their rapid evaporation via Hawking radiation, which likely precludes extended observation. Despite these difficulties, exploring the properties of such black holes may offer valuable insights into the interplay between general relativity and quantum theory, and thus warrants further investigation.

\begin{acknowledgments}
%|--------------------------------------------------------------------|
We would like to thank Wencong Gan, and Chen Lan for helpful discussions. We sincerely appreciate Cong Zhang for valuable suggestions that greatly improved the quality of this work. We also gratefully acknowledge the anonymous referee for their constructive feedback and critical comments, which helped us significantly enhance the clarity and completeness of the manuscript. Yan Liu is financially supported by Natural Science Foundation of Shandong Province under Grants No.ZR2023QA133 and Yantai University under Grants No.WL22B218. 
Jie Jiang is supported by the National Natural Science Foundation of China with Grant No.12205014, the Guangdong Basic and Applied Research Foundation with Grant No.2021A1515110913.
Bing Sun is supported by the National Natural Science Foundation of China under Grants No.12375046 and No.12475046 and Beijing University of Agriculture under Grants No.QJKC-2023032.
%|--------------------------------------------------------------------|
\end{acknowledgments}

\appendix

\section{Details on the calculations}

\subsection{Generic circular motion}
For generic circular geodesic motion, the radial potential satisfies
 \bea
R(r_*)=0,\qquad \frac{\d R(r_*)}{\d r_*}=0.\label{Rcir}
 \eea
By treating $\alpha$ and $M$ as spacetime parameters, these two equations contains three variables: the circular radii $r_*$, the orbital angular momentum $L$ and the orbital energy $E$. It is convenient to analyze the motion in the space of $(E, L)$, therefore, we solve the equations Eq.\eqref{Rcir} above and obtain the conserved energy and angular momentum of the circular motion given by \eqref{Ecir} and \eqref{Lcir}.

The positive property of $L^2$ is decided by the cubic polynomial 
\bea
\beta=r_*^3M-(r_*-2M)(r_*-4M)\alpha^2,\label{noISCO1}
\eea
which implicates that there exists a critical value of the parameter $\alpha=\alpha_c$, beyond which the angular momentum of the circular motion is unreal.

By simply analyzing the structure of the polynomial $\beta$, we can roughly discuss the disappearance of the circular geodesic motion. In the region $2M<r_*<4M$, the angular momentum of the circular motion is always real. In the region $r_*>4M$, the angular momentum of the circular motion is real with the parameter $\alpha<\alpha_c$. However, when $\alpha>\alpha_c$ in the region $r_*>4M$, the angular momentum of the circular motion could be unreal, which implicates that the circular motion is not always allowed in this region, such that there exists some region, although beyond the ICO, in which no circular orbits exist, even the ISCO may not be allowed neither. 

The explicit range of the "no circular" region can be obtained by finding the root of the cubic polynomial $\beta$ in \eqref{noISCO1}. The discriminate of $\beta$ is given by 
\bea
\Delta_3=\frac{4\alpha^4(\alpha^4-432M^4)}{M^2}.
\eea

When $\alpha<2\times3^{3/4}M$, we have $\Delta_3<0$, there exists only one real root for $\beta$ given by
\bea
r_{*1}&=&\frac{\alpha^2}{3M}-\frac{1}{3}(-\frac{\alpha ^6}{M^3}+3 \sqrt{1296 \alpha ^4 M^2-\frac{3 \alpha ^8}{M^2}}+\frac{27 \alpha ^4}{M}\nonumber\\
&&-108 \alpha ^2 M)^{1/3}-\frac{1}{3}(-\frac{\alpha ^6}{M^3}-3 \sqrt{1296 \alpha ^4 M^2-\frac{3 \alpha ^8}{M^2}}\nonumber\\
&&+\frac{27 \alpha ^4}{M}-108 \alpha ^2 M)^{1/3}\leq r_+
\eea
which implicates that outside of the horizon, we always have $\beta\geq 0$, such that the circular motion is always allowed.

When $\alpha=2\times 3^{3/4}M$, we have $\Delta_3=0$, the roots of $\beta$ are given by
\bea
r_{*0}=r_{*2}=2(\sqrt{3}+3)M\approx 9.464M,\label{rs0}\\
r_{*1}=4(2\sqrt{3}-3)M\approx 1.856M,
\eea
which implicate that we always have $\beta\geq 0$ outside of the horizon, therefore, the circular motion is always allowed.

When $\alpha>2\times3^{3/4}M$, we have $\Delta_3>0$, there exists three distinct real roots for $\beta$ given by
\bea
r_*^{(i)}&=&\frac{1}{3}\left(\alpha^2+2\alpha\sqrt{\alpha^2-18M^2}\cos(\zeta+\frac{2\pi i}{3})\right),\nonumber\\
\zeta &=&\frac{1}{3}\arccos(\frac{108M^4-27M^2\alpha^2+\alpha^4}{\alpha\sqrt{(\alpha^2-18M^2)^3}}),\quad i=0,1,2.\nonumber\\
\eea

Based on the analysis below equation \eqref{ri}, we conclude that the critical value of the parameter and the radius of critical circular motion are given by
\bea
\alpha_c=2\times 3^{3/4}M, \qquad r_c=2(\sqrt{3}+3)M,
\eea
when $\alpha>\alpha_c$, the forbidden region for circular motion appears. 

Note that the radii $r=r_*$ satisfies the zero conditions \eqref{Rcir}. Such that the test particle either move as a circular motion with non-zero angular momentum, or levitates at $r=r_*$ with vanishing angular momentum. However, when the parameter $\alpha$ takes the critical value $\alpha=\alpha_c$, the orbital angular momentum is zero. Therefore, critical behavior of the test particle is the second pattern, levitating at $r=r_*$.

\subsection{ISCO}
\begin{figure}
    \centering
    \includegraphics[width=0.95\linewidth]{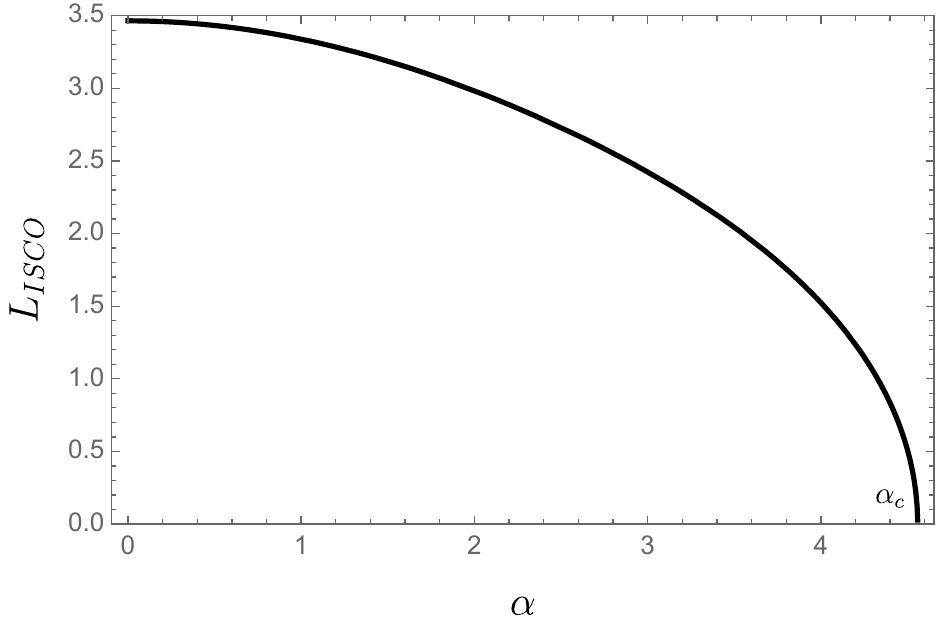}\\
    \includegraphics[width=0.95\linewidth]{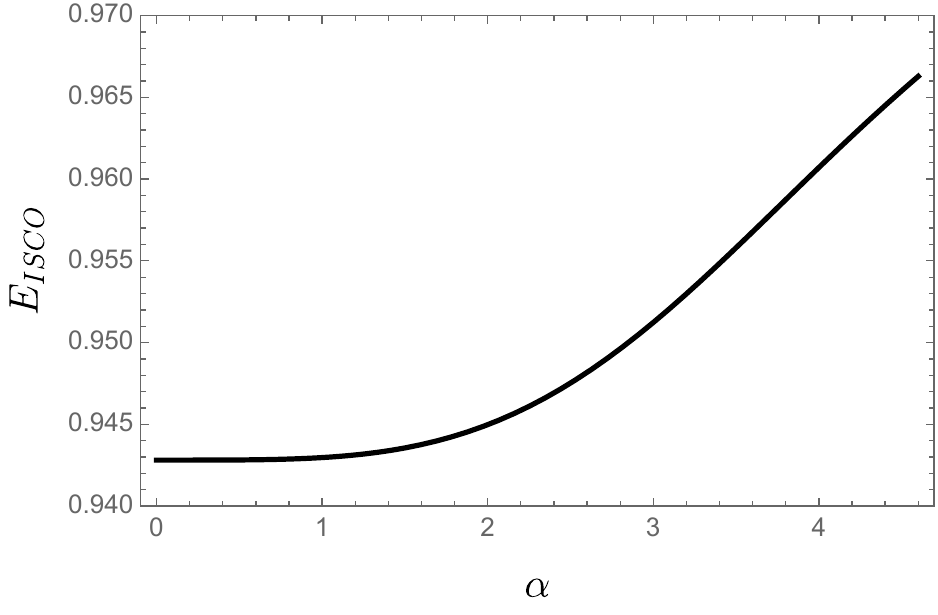}
    \caption{ The variation of the angular momentum and energy of ISCO.}
    \label{fig:ISCO1}
\end{figure}

The position and conserved energy and angular momentum of the ISCO motion can be obtained by solving 
 \bea
R(r)=0,\qquad \frac{\d R(r)}{\d r}=0,\qquad \frac{\d^2 R(r)}{dr^2}=0\label{Rcir1}
 \eea
However, in the end the equations \eqref{Rcir1} turn to be a seventh order polynomial of $r$. Therefore, we solve for the $r_{ISCO}$ numerically, as shown in Figure. \ref{fig:rISCO1}. Then the orbital angular momentum and energy of ISCO can be numerically obtained as well. In Figure \ref{fig:ISCO1}, we plot the variation of conserved energy and angular momentum of ISCO due to the increasing quantum correction $\alpha$.  Figure \ref{fig:ISCO1} shows that as the parameter $\alpha$ increases, while the energy of the ISCO monotonically increases, the angular momentum of ISCO monotonically decreases to zero and ends at $\alpha=\alpha_c$.

When $\alpha$ is small, the position, conserved energy, and angular momentum of the ISCO due to quantum correction are given by
\bea
&&r_{ISCO}=6M+\frac{\alpha^4}{81M^3}+O(\alpha^6),\\
&&L_{ISCO}=2\sqrt{3}M-\frac{2\alpha^2}{9\sqrt{3}M}+O(\alpha^4),\\
&&E_{ISCO}=\frac{2\sqrt{2}}{3}+\frac{\alpha^4}{4374\sqrt{2}M^4}+O(\alpha^6).
\eea

\subsection{Marginal circular motion}
The marginal circular orbit is the circular orbit with the energy $E=1$. By solving \eqref{Rcir},  the angular momentum and the location of the MCO motion $L_{m}$ and $r_{m}$ should satisfy
\bea
L_{m}&=&\sqrt{\frac{r^2 ( M (r^3 + 6 r \alpha^2)-8 M^2 \alpha^2 - r^2 \alpha^2 )}{(r-3 M ) (r^3 - 4 M \alpha^2 + 2 r \alpha^2)}},\label{Lm}\\
0&=&r^4 (-3 M + r) (r^3 - 4 M \alpha^2 + 2 r \alpha^2)\nonumber\\
&&-(-2 M + r)^2 (r^3 - 2 M \alpha^2 + r \alpha^2)^2. \label{rm}
\eea

\begin{figure}
    \centering
    \includegraphics[width=0.95\linewidth]{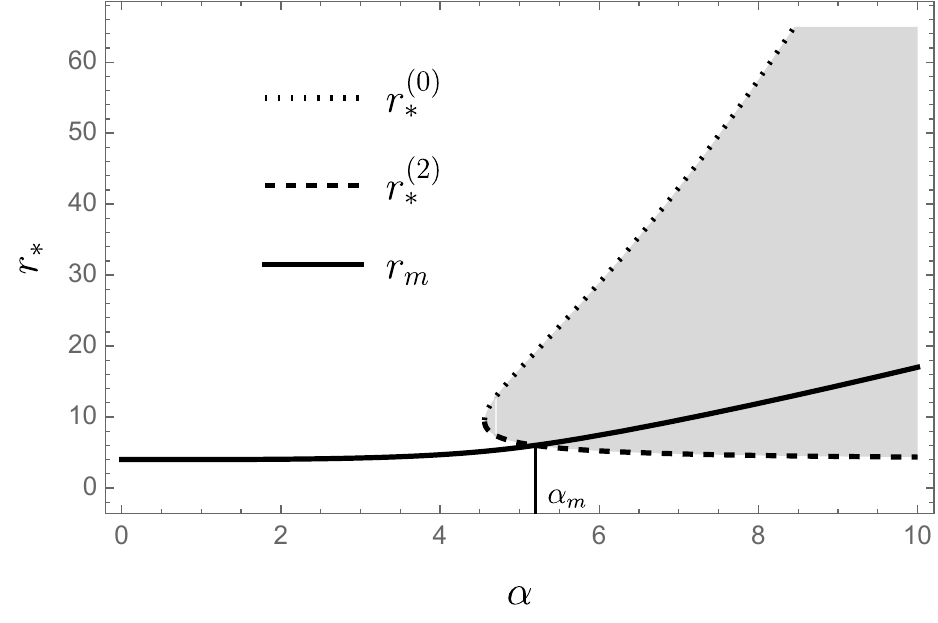}\\
    \includegraphics[width=0.95\linewidth]{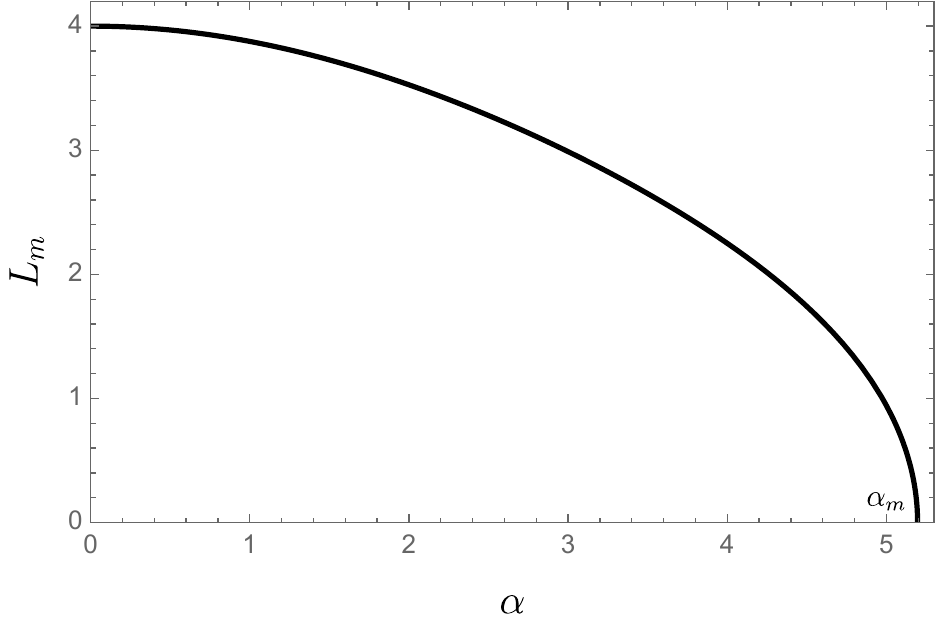}
    \caption{The disappearance and the angular momentum of MCO.}
    \label{fig:rLmbo}
\end{figure}

Note that even though the expression of the $L_m$ in Eq.\eqref{Lm} has same form with the generic circular case in Eq. \eqref{Lcir}, but MCO does not disappear at $\alpha_c$. In Figure \ref{fig:rLmbo} we show the disappearance and angular momentum of MCO motion due to the loop quantum correction $\alpha$. The line denoting MCO enter the gray region after intersecting with $r_*^{(2)}$ at $\alpha=\alpha_m$. The explicit critical value for MCO can be obtained by solving $r_m=r_*^{(2)}$ and given by
\bea
\alpha_m=3\sqrt{3}M, \qquad r_{mc}=6M.
\eea
When $\alpha>\alpha_m$, MCO does not exist. Similarly with ISCO, the angular momentum of MCO monotonically decrease to zero with increasing $\alpha$, and ends at $\alpha=\alpha_m$.

When $\alpha$ is small, the position and angular momentum of the MCO due to the quantum correction are given by
\bea
&&r_{m}=4M+\frac{\alpha^4}{256M^3}+O(\alpha^6),\\
&&L_{m}=4M-\frac{\alpha^2}{8M}+O(\alpha^4).
\eea

\subsection{Bound motion}

The turning points of the bounded motion are the pericenter $r_p$ and apocenter $r_a$, satisfying 
\bea
R(r_a)=R(r_p)=0,\label{Rbound}
\eea
replacing $r_a$ and $r_p$ by the taxonomy of eccentricity $e$ and semi-latus rectum $p$ as
\bea
r_a=\frac{p}{1-e},\qquad r_p=\frac{p}{1+e},
\eea
and solve \eqref{Rbound}, the energy and angular momentum of the bound motion are obtained given by
\bea
E^2&=&\frac{1}{p^6}(2 (1 + e) M - p) ((1 + e)^2 L^2 + p^2) \nonumber\\
&&(  (2 (1 + e) M - p)(1 + e)^2 \alpha^2-p^3 ),\\
L^2&=&\frac{p^2(M p^3 - (4 M - p) (2 (1 + e^2) M - p) \alpha^2)}{(p-((3 + e^2) M) ) (p^3 + 
   2 (p + e^2 p-2 M - 6 e^2 M) \alpha^2)}.\nonumber\\
\eea
The angular momentum of bound orbits are real when $\alpha\leq \alpha_{bc}$, where the critical parameter $\alpha_{bc}$ for the bound motion is given by
\bea
\alpha_{bc}=\sqrt{\frac{Mp^3}{(4 M - p) (2M (1 + e^2)- p)}}.
\eea

\section{From plunging orbits to deflecting orbits}\label{appplunge}
\begin{figure}
    \centering
    \includegraphics[width=0.9\linewidth]{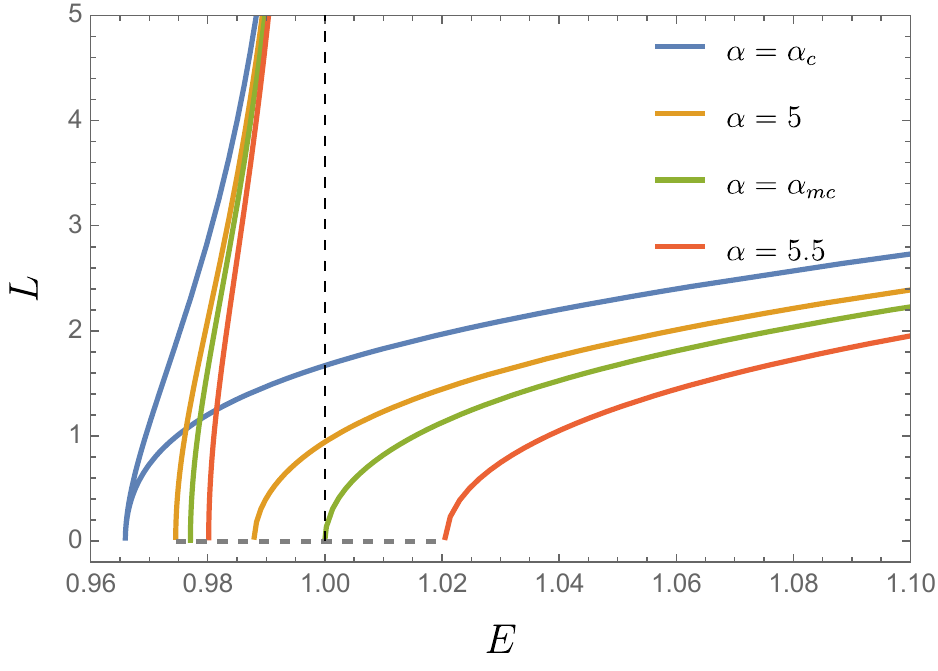}
    \caption{The variation of the "no-circular" region presented by the gray dashed line.}
    \label{fig:LEplunge}
\end{figure}
For the purely plunging orbits with structure $\vert +\rangle$, the radial potential satisfies $R(r,E,L)>0$ outside of the horizon. In the Schwarzschild spacetime, as shown in Figure \ref{fig:SchwarzEL}, such plunging orbits exists in the region $E\leq 1$ and $L<L_{cir}$, where $L_{cir}$ denotes the angular momentum of the circular orbit. In Figure~\ref{fig:LEplunge}, we extend the phase space to larger $\alpha$ values to illustrate how the critical region varies relative to Figure \ref{fig:CIrEL}. In particular, when $\alpha>\alpha_{mc}$, the the phase space structure changes markedly. Physically, the points on the gray dashed line represent the critical motion with $L=0$. By continuity, the root structures of these critical points are the same as those of their neighboring regions. 

For example, in Figure~\ref{fig:LEplunge} the red curve with $\alpha=5.5$ shows that the right segment with $E<1$ corresponds to unstable circular motion, which we label $E_u$ for convenience. The critical points on the gray dashed line between $E=1$ and $E_u$ thus represent deflecting orbits within $\vert+\bullet-\bullet+\rangle$. By contrast, for the green curve with $\alpha=\alpha_{mc}$, these points correspond to purely plunging orbits $\vert+\rangle$, as can be verified by comparing the root structure distributions in Figure~\ref{fig:SchwarzEL}. This contrast indicates that quantum effects can alter the orbital behavior, transforming plunging motion into deflecting motion. 

\begin{figure}
    \centering
    \includegraphics[width=0.95\linewidth]{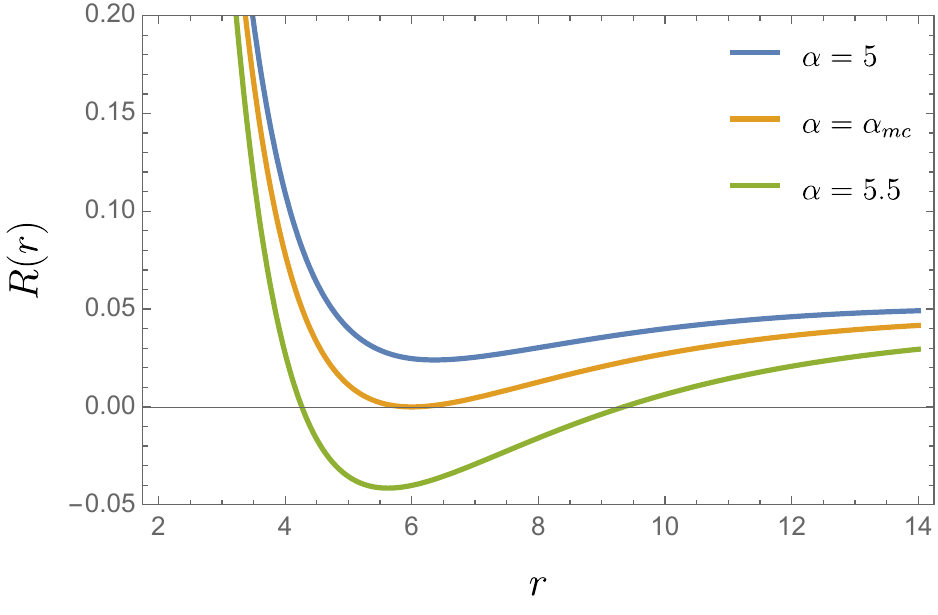}
    \caption{The variation of the radial potential $R(r)$ with $E=1$ and $L=0$.}
    \label{fig:Rplunge}
\end{figure}

The change is even clearer when examining the radial potential directly, as shown in Figure \ref{fig:Rplunge}. For specific orbital energy and angular momentum with $\alpha<\alpha_{mc}$, the radial potential remains positive outside the horizon, allowing the particle to plunge directly into the black hole. As $\alpha$ increases beyond $\alpha_{mc}$, a turning point emerges and the originally plunging orbit becomes a deflecting one. Because the angular momentum remains zero, this motion is a critical deflecting motion, thus the particle arrives from infinity and bounce back along the same path. However, it is worth noting that in a universe with fixed $\alpha$, the purely plunging motion $\vert+\rangle$ does not exhibit repulsive behavior if it's orbital energy is sufficiently large to overcome the effective repulsive force, which will be explained in Appendix \ref{appforces}.

\section{Repulsive forces}\label{appforces}
The attractive or repulsive force that a particle experienced in a curved spacetime can be treated as the force required to hold a particle static at a spacial position $r$. The 4-velocity of this particle is aligned with the timelike Killing vector $\xi^\mu=\p_t$. The 4-acceleration of this particle is defined as 
\bea
a^\mu=u^\nu\nabla_\nu u^\mu,
\eea
where $\nabla_\mu$ is the covariant derivative, $u^\mu=(1/\sqrt{f(r)},0,0,0)$ is the 4-velocity of this static particle, and $f(r)$ is defined in \eqref{fr}. Thus the force experienced by a unit–mass particle is given by  
\bea
f_r=-\frac{\beta}{r^{5}}\label{force}
\eea
where $\beta$ is defined in Eq.~\eqref{noISCO}. This expression shows that the repulsive component of the force completely originates from the polynomial $\beta$, i.e., from the quantum parameter $\alpha$.

\begin{figure}
    \centering
    \includegraphics[width=0.9\linewidth]{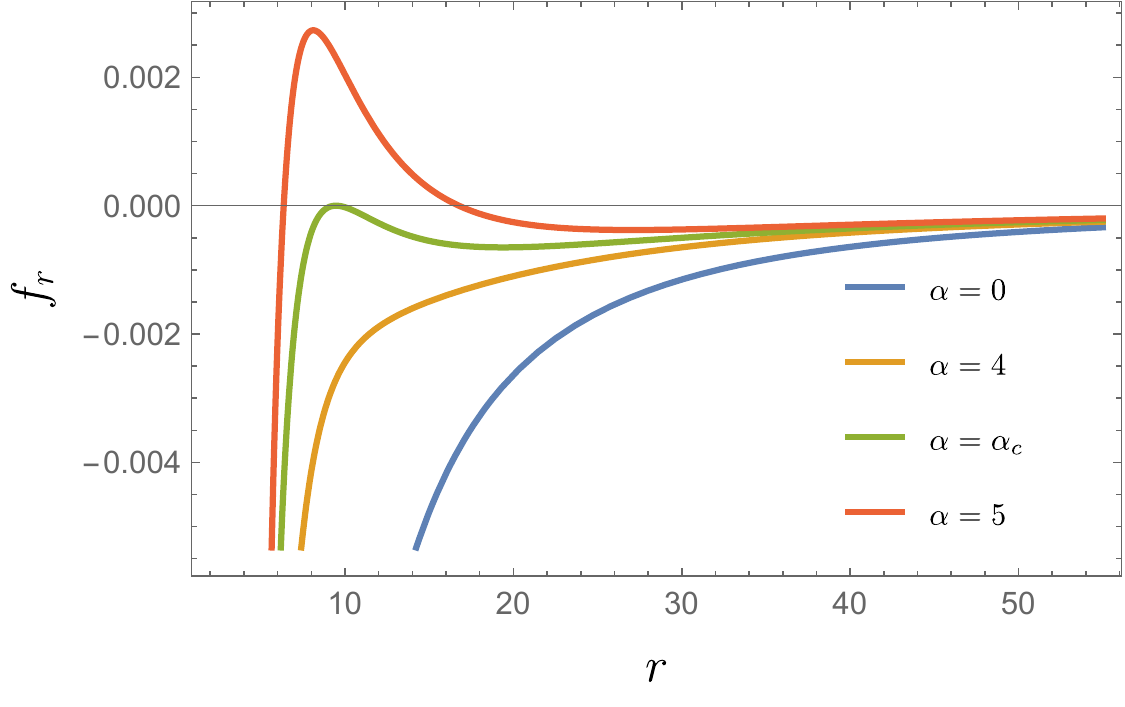}
    \caption{The variation of the radial force $f_r(r)$ with values of $\alpha$.}
    \label{fig:fr}
\end{figure}

As illustrated in Fig.~\ref{fig:fr}, when $\alpha < \alpha_c$ the radial force $f_r(r)$ is purely attractive, whereas for $\alpha > \alpha_c$, a repulsive region emerges.  
The extent of this repulsive region grows with increasing $\alpha$.  
Note that the orbital energy $E$ affects the behavior of the geodesic motion. When $E$ is large enough, the particle could overcome the repulsive force and still show plunging behavior shooting into the black hole.
This explains why, even when a repulsive component exists for $\alpha > \alpha_c$, certain geodesic motions may not exhibit repulsive behavior.

\bibliography{refs}
\bibliographystyle{utphys}

\end{document}